\documentclass[conference]{IEEEtran} 
\IEEEoverridecommandlockouts
\usepackage{cite}
\usepackage{amsmath,amssymb,amsfonts}
\usepackage{algorithmic}
\usepackage{graphicx}
\usepackage{textcomp}
\usepackage{xcolor}
\usepackage{tikz}
\usepackage{latexsym}
\usepackage{subcaption}
\usepackage{hyperref}

\usepackage[a4paper, total={184mm,239mm}]{geometry}
\def\BibTeX{{\rm B\kern-.05em{\sc i\kern-.025em b}\kern-.08em
    T\kern-.1667em\lower.7ex\hbox{E}\kern-.125emX}}
\begin{document}

\title{Design of High-Throughput Mixed-Precision CNN Accelerators on FPGA}

\author{\IEEEauthorblockN{Cecilia Latotzke, Tim Ciesielski, and Tobias Gemmeke}
\IEEEauthorblockA{
Chair of Integrated Digital Systems and Circuit Design, RWTH Aachen University, 52062 Aachen Germany\\
Email: \{latotzke, gemmeke\}@ids.rwth-aachen.de, tim.ciesielski@rwth-aachen.de
\thanks{Accepted at 32nd International Conference on Field Programmable Logic and Applications (FPL 2022)}
}
}

\maketitle
\begin{abstract}
Convolutional Neural Networks (CNNs) reach high accuracies in various application domains, but require large amounts of computation and incur costly data movements. One method to decrease these costs while trading accuracy is weight and/or activation word-length reduction. Thereby, layer-wise mixed-precision quantization allows for more efficient results while inflating the design space. In this work, we present an in-depth quantitative methodology to efficiently explore the design space considering the limited hardware resources of a given FPGA. Our holistic exploration approach vertically traverses the various design entry levels from the architectural down to the logic level, and laterally covers optimization from processing elements to dataflow for an efficient mixed-precision CNN accelerator. Our resulting hardware accelerators implement truly mixed-precision operations that enable efficient execution of layer-wise and channel-wise quantized CNNs. Mapping feed-forward and identity-shortcut-connection mixed-precision CNNs result in competitive accuracy-throughout trade-offs: 245 frames/s with 87.48\% Top-5 accuracy for ResNet-18 and 92.9\% Top-5 accuracy with 1.13 TOps/s for ResNet-152, respectively. Thereby, the required memory footprint for parameters is reduced by 4.9$\times$ and 9.4$\times$ compared to the respective floating-point baseline.
\end{abstract}


\maketitle

\section{Introduction}
Convolutional Neural Networks (CNN) excel in classification and detection tasks. 
Some tasks, like autonomous driving, require rapid decision making, hence, latency and throughput of CNNs have to be optimized. 
To bring the classification closer to the data, CNNs are accelerated on edge devices.
But the CNN's high performance in quality-of-service comes with high computational and data transfer cost \cite{Horowitz2014}.
On edge devices with a limited energy budget the relevance of this cost is even more pronounced.
Hence, the execution of CNNs on edge devices requires optimization steps in terms of CNN model and supportive hardware.

Reduced precision provides one way to reduce the computational and data transfer cost triggering a transition from 32\thinspace bit floating-point to limited fixed-point (e.g., 8 bit going even down to binary operands \cite{Stadtmann2020}) representation in the inference of CNNs.
So the memory footprint can be reduced to a quarter or less than its original size. 
Even though the goal is to reduce the cost for processing CNNs, the quality-of-service should be preserved.
For this work the quality-of-service is accuracy, as our application scenario is image classification, specifically the dataset ImageNet \cite{ImageNet}.
It is commonly accepted that 8\thinspace bit fixed-point is sufficient to maintain accuracy \cite{Sze2019}.
However, recent publications indicate a trend towards non-symmetric word-lengths of activations and weights that offer accuracy gains compared to their floating-point or symmetric word-length counterparts \cite{Yang2019, Wu2018, Wang2018, Wang2020, Esser2019, Cho2020}.
This effect is attributed to the induced quantization noise having a regularizing effect on the performance of the CNN \cite{Esser2019}. 
In the end, the optimal choice of word-length per layer will remain a topic of research while being subject to the adopted CNN model and application-dependent accuracy requirements \cite{Lin2016}.
Hence, edge devices will certainly benefit from hardware architectures that efficiently translate any kind of word-length reductions to improvements in efficiency.

Today, hardware accelerators for CNNs are mostly optimized for a specific word-length or even a specific CNN \cite{Latotzke2021, Mittal2020}.
Designing hardware accelerators based on FPGAs actually offers, thanks to their reconfigurability, the opportunity to use a hardware mapping optimized specifically for the type of CNN under consideration.
However, this opportunity spans an extensive design space which has not been systematically evaluated combining all levels from realization of processing element (PE) up to the architectural level.
To close this gap, this work presents the first holistic Design Space Exploration (DSE) for mixed-precision accelerators targeting efficient support of varying word-lengths.

The main contributions of this paper are the following.
\begin{itemize}
    \item Our work provides a proportionate increase in throughput with word-length reduction for mixed-precision CNNs while preserving flexibility to support variable word-lengths and CNN models on an FPGA.
    \item  Our holistic DSE combines all three, the PE realization, the PE array dimensions, and the dataflow, to maximize the throughput of feedforward CNNs as well as identity-shortcut-connected CNNs.
    \item We assess in detail partial product processing by means of a quantitative evaluation of Multiply-Accumulate (MAC) PE architectures that support mixed-precision by segmenting the PE into multiple Partial Product Generators (PPG) and an adder tree. 
    With the segmentation, the word-length of weights can be adjusted on-the-fly enabling layer-wise or channel-wise quantization without changing the FPGA image.
    \item We benchmark embedded DSP macros vs. FPGA's logic fabric in terms of consumed energy when executing mixed-precision arithmetic.
    \item Finally, we apply our holistic DSE methodology to different mixed-precision ResNet models \cite{ResNet} and evaluate the resulting set of specific hardware accelerators.
\end{itemize}

The paper is organized as follows.
In Section\thinspace\ref{sec:RelatedWork}, an overview of the related work is given.
The DSE in Section\thinspace\ref{sec:DSE} is divided into the design dimensions on the PE level in Part A and on the CNN accelerator dataflow in Part B.
The results of this DSE are presented in Section\thinspace\ref{sec:results}.
Here, the accuracy-throughput trade-off for mixed-precision CNNs on the created designs are presented including a comparison to the state-of-the-art.
Section\thinspace\ref{sec:Conclusion} concludes this work.

\section{Related Work}
\label{sec:RelatedWork}
There is an extensive body of work addressing the design of optimized hardware accelerators for CNNs.
Most previous works limit their DSE to the dataflow optimization by either considering the memory hierarchy or options to unroll or tile loops.
Some works extend this design space to include the PE array dimensions \cite{Tsimpourlas2018,Parashar2019,Yang2020,Ma2018,Rahman2017,Reggiani2019,Zhong2017, Lu2017,Veneries2017,Kwon2020,Blott2018,Samajadar2019}.
But there is only a small number of previous works, which explore the design space for dataflow for mixed-precision CNN accelerator implementations on FPGAs \cite{Nguyen2021,Blott2018}.
{In \cite{Blott2018}}, two  different dataflows of either storing and processing all weights of a small artificial neural network at once on an FPGA or storing and processing only one layer of a bigger CNN on an FPGA and then processing and storing the next layer of this CNN are presented.
The considered PEs support weights and activations with 1, 2, and 3\thinspace bit. 
In \cite{Nguyen2021} the CNN layers are grouped according to their dominant memory access pattern either into DRAM dominated (for fully connected layers FC) or BRAM dominated (for convolutional layers CONV).
To increase the efficiency of CONV layers, most weights are binarized, and only a few weights are kept at 8\thinspace bit.

\begin{figure}[htb]
\begin{center}
\includegraphics[scale= 0.65]{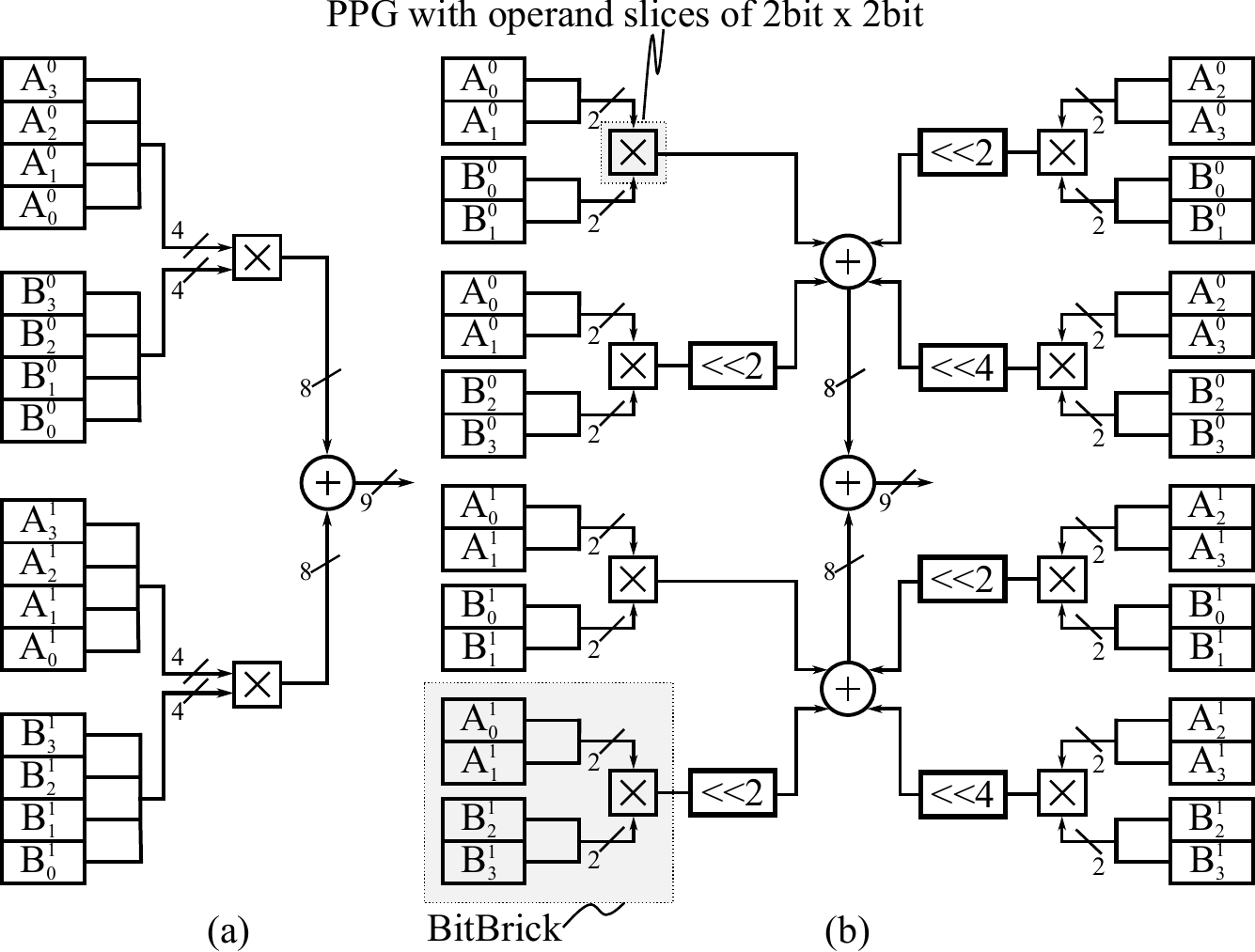}
\caption{Schematic with an input word-length of 4\thinspace bit and (a) a conventional MAC PE and (b) a MAC PE with PPG of a size of 2\thinspace bit$\thinspace\times\thinspace$2\thinspace bit according to \cite{Sharma2018,Kim2019}.}
 \label{fig:bitbrick}  
\end{center}
\vspace{-0.5cm}
\end{figure}

{An exploration combining the optimization on PE level considering the intra-PE level up to inter-PE level (the PE array) is found in \cite{Zhong2017}, but focusing only on single-precision 32\thinspace bit floating-point representation.}
We define a PE as an entity to facilitate MAC operations, hence, it consists of a multiply and an accumulate unit.
Its multiply unit can be sliced into multiple PPGs (cf. Fig.\thinspace\ref{fig:bitbrick}), with each operand slice defining how many bits of the input word can be processed in one PPG.
Hence, one PE comprises multiple partial products.
If the input word-length is equal to the operand slice, each PPG processes a different word.
{If the input word-length is bigger than the operand slice, it utilizes multiple PPGs, cf. \thinspace Fig.\thinspace\ref{fig:bitbrick}\thinspace b.}
And if the input word-length is smaller than the operand slice, a part of the PPG stays idle.
So far, such 2\thinspace bit\thinspace$\times$\thinspace2\thinspace bit PPGs were introduced as BitBricks in BitFusion \cite{Sharma2018}.
BitBlade \cite{Ryu2019} extended BitFusion by exploring the array architecture composed of such 2\thinspace bit\thinspace$\times$\thinspace2\thinspace bit PPGs. 
Finally, the MAC unit review \cite{Camus2019} utilized also only PPGs with operand slices of 2\thinspace bit.
In contrast to referenced research, we consider the size of the operand slice as an explicit parameter in our DSE.

In conclusion, to the best of our knowledge, there is currently no DSE published considering the impact of all levels from intra-PE to architectural on the efficiency of the overall CNN accelerator.
Our work uses the taxonomy introduced in \cite{Camus2019} as baseline, which groups PE designs of previous works with regards to their way of processing inputs and internally consolidating the computed results.
The work of \cite{Camus2019} focussed on the PE level, only.
Our work extends the DSE by adding a system-level perspective.
{The presented work firstly combines for different word-lengths the exploration of PE realization, PE array dimensions, and dataflow optimization in one holistic DSE.}

\section{Design Space Exploration Methodology}
\label{sec:DSE}

In this work, the examined design space is spanned by the following four dimensions: (1) the PE design options with regards to their input processing and data consolidation styles and the different operand slices of parallel multiplication units (PPGs) in a PE, (2) the total number of PEs, (3) the PE array dimensions, (4) and the dataflow optimizations (i.e., the temporal and spatial data reuse).
\begin{figure}[htb]
\begin{center}
\includegraphics[scale= 0.75]{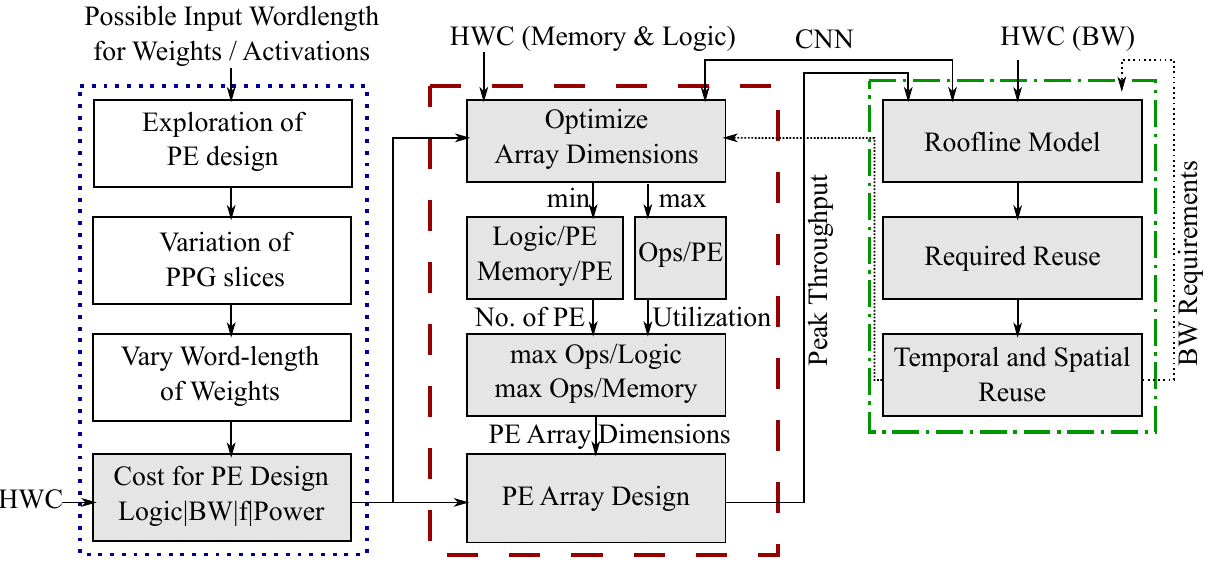}
\caption{Flowchart for DSE for the PE (blue dotted box), the PE array (red dashed box), and the dataflow (green dashed-dotted box), with hardware constraints (HWC) and bandwidth (BW).}
\label{fig:DSE_flow_Chart}  
\end{center}

\end{figure}

The constraints and resources of our DSE are geared towards FPGAs and comprise the selected set of supported input word-lengths, the chosen mixed-precision CNNs, and the hardware constraints including the available logic, memory, and bandwidth.
The specific hardware constraints considered in this paper are set by the choice of a Stratix\thinspace V FPGA.
In any case, the presented DSE methodology can generically be applied to any FPGA architecture.

The DSE comprises three key phases with fully automated blocks highlighted in gray (cf. Fig.\thinspace\ref{fig:DSE_flow_Chart}).
These phases are the semi-automatic PE DSE, the fully automated PE array DSE, and the fully automated system-level evaluation of the total dataflow.
Furthermore, selected highly regular structures were handcrafted for best efficiency.
Key interdependencies in the DSE are indicated with dotted lines.
We focus in this work on the processing of CONV layers because of their dominant contribution to total throughput and energy \cite{Blott2018}.

\subsection{Precision Scalable Processing Element}

\begin{figure}[htb]
\begin{center}
\includegraphics[scale= 0.65]{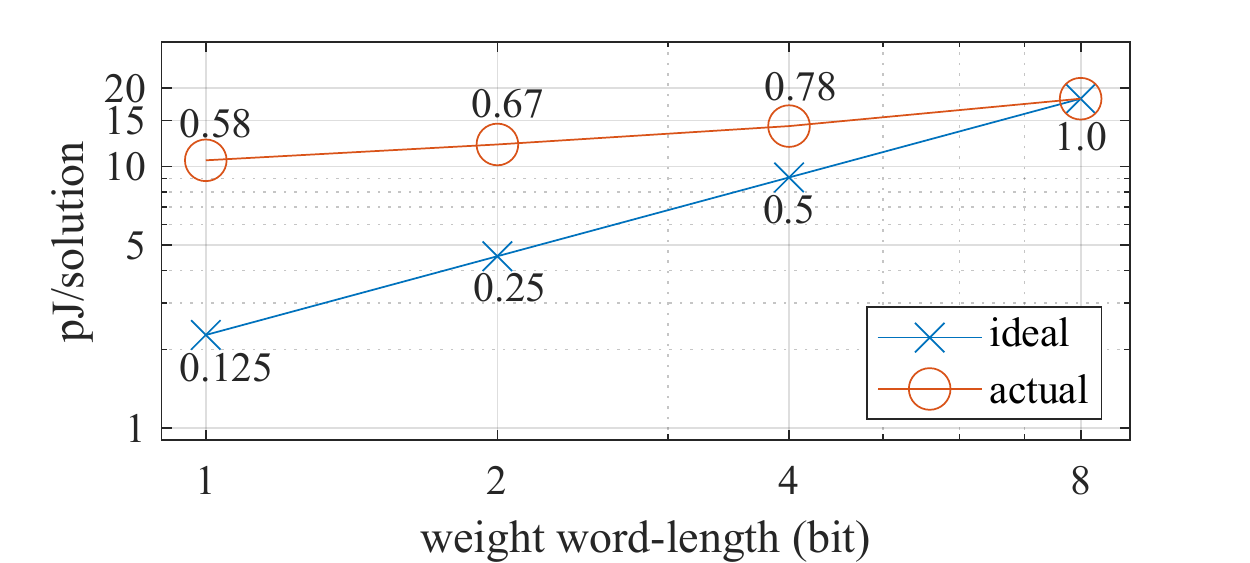}
\caption{Energy of multiplication for Stratix\thinspace IV DSPs, with activations always 8\thinspace bit and varying word-length for weights. Solution stands here for MAC product.}
 \label{fig:DSP_energy}  
\end{center}
\vspace{-0.2cm}
\end{figure}

The typical PE design of a MAC unit uses dedicated DSP hardmacros of the FPGA to efficiently process the MAC operations.
The highest efficiency is achieved by exploiting the full word-lengths of such hardmacros.
However, energy reduction does not scale linearly with the reduction in word-length as exemplified in Fig.\thinspace\ref{fig:DSP_energy} and indicated by comparing linear scaling with actual 
numbers. A word-length reduction from 8 to 1\thinspace bit only provides a 0.58$\times$ energy reduction instead of the ideal reduction of 0.125$\times$. Furthermore, exclusively relying on DSPs limits the degree of parallelism to the number of available hardmacros on a specific FPGA.

\begin{figure}[htb]
\begin{center}
\includegraphics[scale= 0.65]{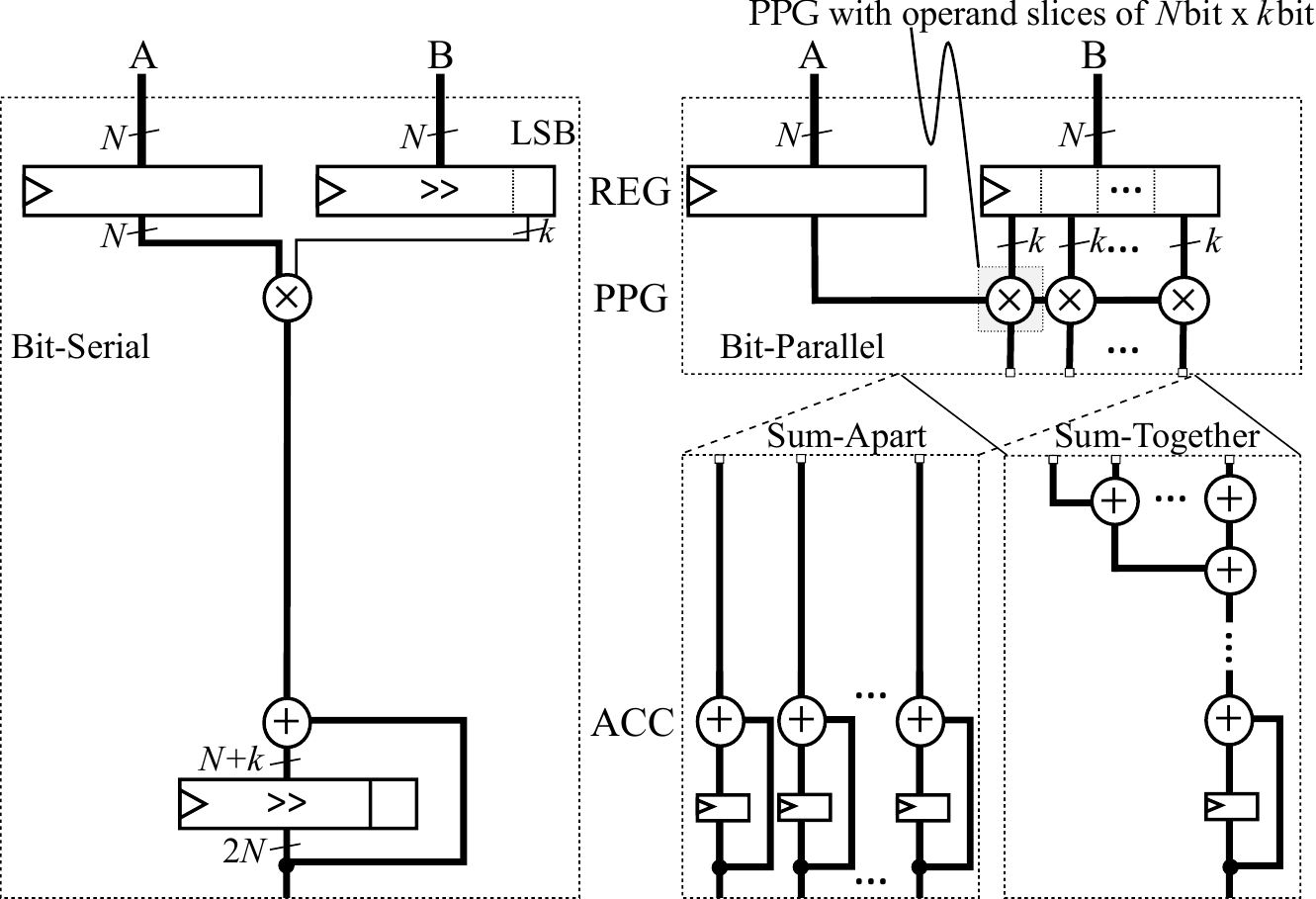}
\caption{Schematics of different PE design options for the case of 1D scaling.
The processing of inputs in Bit-Serial or Bit-Parallel fashion as well as the data consolidation of the partial sums by means of an adder tree, Sum-Apart, or its absence, Sum-Together, are shown.}
 \label{fig:CamusTaxonomy}  
\end{center}
\vspace{-0.5cm}
\end{figure}

The design space of PEs is divided into four dimensions.
One dimension determines whether the input is processed in serial or in parallel.
Another dimension reflects the operand slice of the PPG in the PE for the parallel case or the bits\thinspace/\thinspace cycle for the serial case. 
The third dimension defines the option to scale either both inputs, 2D (cf. Fig.\thinspace\ref{fig:bitbrick}) or only one input, 1D (cf. Fig.\thinspace\ref{fig:CamusTaxonomy}).
For the 2D case the operand slice is $k$\thinspace bit\thinspace$\times\thinspace k$\thinspace bit
and for the 1D case the operand slice is $N$\thinspace bit\thinspace$\times\thinspace k$\thinspace bit. With $N$ being the input word-length, only $k$ is reported as operand slice.
Together they determine how many bits of the input are multiplied in each partial product.
And finally, the fourth dimension describes the data flow of the accumulation, whether the partial sums are accumulated in an adder tree or stored individually.
Fig.\thinspace\ref{fig:CamusTaxonomy} visualizes this generic principle for the 1D case, 
specific schematics of many variants can be found in \cite{Camus2019}.
Bit-Serial (BS) processing in time with $k$\thinspace bit\thinspace/\thinspace cycle is shown in Fig.\thinspace\ref{fig:CamusTaxonomy} (left), for $k$ equal to 1 the multiplication is only an AND operation.
Bit-Parallel (BP) processing on word-level is shown in Fig.\thinspace\ref{fig:CamusTaxonomy} (right) with the $N$-bit input bus split into $N$\thinspace/\thinspace$k$ slices of $k$ bits. 
Accumulation is done independently per partial product (cf. Sum-Apart (SA) in Fig.\thinspace\ref{fig:CamusTaxonomy} bottom left) or across all partial products using a preceding adder tree (cf. Sum-Together (ST) in Fig.\thinspace\ref{fig:CamusTaxonomy} bottom right). 
These different PE design options can be mixed and matched with various PE designs.

\subsection{Dataflow}

Dataflow comprises data reuse, data access, and data storage patterns.
An optimal dataflow maximizes the computational intensity \cite{Blott2018,Williams2009} by either broadcasting data to multiple parallel working PEs or by reusing data over time.
Temporal reuse requires additional registers, whereas spatial reuse requires additional routing.
With a decrease in spatial reuse the total throughput declines, therefore, we focus on the maximization of spatial reuse.

The spatial reuse comprises the reuse of weights, the reuse of activations, and the reuse of partial sums in the PE array.  
The reuse factor of each of them depends on tiling of the convolutional loops along three dimensions: the input feature map height $I_\text{H}$, the input channel width $I_\text{W}$, and the output channel depth $O_\text{D}$ (cf. Table\thinspace \ref{tab:reuse}).
These dimensions differ in a given CNN per layer. 
Since the PE array dimensions {heigth $H$, width $W$, and depth $D$} are fixed for a given hardware accelerator design they have to be optimally chosen according to the CNN and the hardware constraints.
The chosen PE array dimensions determine on the one hand, the total number of PEs ($N_\text{PE}$) in the PE array expressed in Eq.\thinspace\ref{eq:NPE}, and on the other hand, the total spatial reuse, i.e., the total number of parallel accessed BRAMs ($BRAM_\text{NPA}$), cf. Eq.\thinspace\ref{eq:reuse}, With $N$ being the word-length for activations and $w_\text{Q}$ being the word-lengths for weights.
The factor \begin{math}\frac{N}{w_\text{Q}}\end{math} influences the $BRAM_\text{NPA}$ for activations.
\begin{equation}
\begin{split}
N_\text{PE} = {H}\times {W}\times {D}
\end{split}
\label{eq:NPE}
\end{equation}
\begin{equation}
\begin{split}
BRAM_\text{NPA} = \underbrace{{H}\times {D}}_{BRAM_\text{partial sums}} + \underbrace{{H}\times {W}\times\frac{N}{w_\text{Q}}}_{BRAM_\text{activations}} + \underbrace{{W}\times{D}}_{BRAM_\text{weights}}
\\\ \text{with }w_\text{Q} \geq k
\end{split}
\label{eq:reuse}
\end{equation}
\begin{equation}
\begin{split}
U(l) =\frac{P_\text{ideal}(l)}{P_\text{actual}(l)} =\frac{\frac{I_\text{H}^2\times I_\text{W}\times O_\text{D}\times(\frac{K}{S})^2}{{H}\times{W}\times\frac{{N}}{w_\text{Q}}\times D}}{\lceil\frac{I_\text{H}}{H}\rceil\times\lceil\frac{I_\text{W}}{W\times\frac{{N}}{w_\text{Q}}}\rceil\times\lceil\frac{O_\text{D}}{D}\rceil\times I_\text{H}\times(\frac{K}{S})^2}
\end{split}
\label{eq:util}
\end{equation}

Temporal and spatial reuse form together the total reuse of data, which is needed to process a CNN on a given hardware.
The ideal temporal reuse $P_\text{ideal}$ per layer $l$ and actual temporal reuse $P_\text{actual}$ determine the utilization per layer $U(l)$, cf. Eq.\thinspace\ref{eq:util} with filter kernel $K$, stride $S$, and the ceil function $\lceil\cdot\rceil$.
Furthermore, the temporal reuse $P_\text{actual}$ defines the required bandwidth, which is fed back to the roofline model (cf. Fig.\thinspace\ref{fig:DSE_flow_Chart} green box).
This assures that the bandwidth limitations in the different levels of the memory hierarchy are met.
The overall dataflow is automatically adapted with regards to the maximum throughput of the design.
Hence, the dataflow is majorly influenced by the design of the PE array.
\begin{table}[htbp]
\caption{Spatial reuse for unrolling}
\begin{center}
\begin{tabular}{|c|c|c|}
\hline
\textbf{PE array dimension} &\textbf{reuse}&  \textbf{no reuse} \\
\hline
\hline
${H}$& weights& activations, partial sums   \\
\hline
${W}$& partial sums& weights, activations   \\
\hline
${D}$& activations& weights, partial sums   \\
\hline
\end{tabular}
\label{tab:reuse}
\end{center}
\end{table}

The PE array dimensions are optimized with regards to maximizing throughput, i.e., minimizing $P_\text{actual}$, maximizing $U(l)$, and maximizing the operations per FPGA resources (cf. Fig.\thinspace\ref{fig:DSE_flow_Chart} red box). 
The relevant FPGA resources are the available memory resources, BRAMs, and the available logic, lookup tables (LUTs).
The greedy optimization approach for the PE array dimensions explores all possible solutions for a certain mixed-precision CNN, PE design, and hardware constraints. These solution are then compiled to evaluate their feasibility on the chosen FPGA by means of its compilation tool.

\begin{figure}[htb]
\begin{center}
\includegraphics[scale = 0.6]{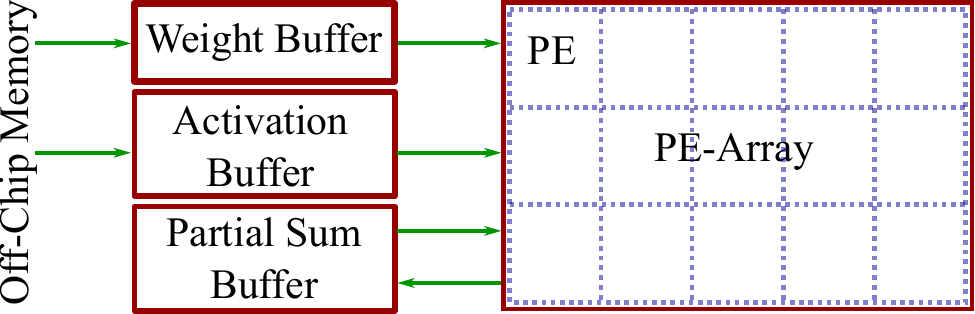}
\caption{The general mixed-precision CNN accelerator scheme. 
The colors correlate with the design space exploration in Fig.\thinspace\ref{fig:DSE_flow_Chart}.}
\label{fig:dataflowScheme}
\end{center}
\vspace{-0.5cm}
\end{figure}

To maximize the throughput, the computational resources are maximized, hence, all available logic is used for computation.
This is realized by choosing a flat memory hierarchy. The on-chip memory is divided in three global buffers with their size based on Eq.\thinspace\ref{eq:reuse} to efficiently broadcast the currently required weights, activations, and partial sums into the PE array. All images, which have to be classified, as well as weights and biases of the accelerated CNN are stored in the off-chip memory and transferred only once to the on-chip memory.
The general design scheme of the mixed-precision CNN accelerator is shown in Fig.\thinspace\ref{fig:dataflowScheme}. 
The PE array is depicted in 2D, although, in reality the optimal PE arrays are always three dimensional with different width $W$, height $H$, and depth $D$ depending on the mixed-precision CNN dimensions.

\section{Results}
\label{sec:results}

Our DSE focuses on the utilization of the available logic of Stratix\thinspace V
leaving a heterogeneous mapping including DSPs for future work.
As one drawback, there is currently no specific gate-level timing simulation support for the Stratix\thinspace V. 
Hence, our DSE relies on scaled Stratix\thinspace IV results to close this gap. Energy numbers are extracted from gate-level timing analysis using activity files assuming uniformly distributed input vectors.

\subsection{Design Space Exploration of Processing Elements}

To start the PE DSE, objectives and boundary conditions have to be defined as follows.
First, the design space is divided into whether both inputs or only one input should offer flexible word-length, i.e., a 1D or 2D scalable design. 
Second, either area or latency can be minimized by choosing BS or BP processing, respectively.
Because a BS design as in Fig.\thinspace\ref{fig:CamusTaxonomy} minimizes the required area per PE while reducing the throughput per PE.
This principle is reversed for the BP design.
Third, the minimum number of parallel processable bits has to be set. 
In case of a BP design with different input word-lengths as operand slice $k$, either the utilization of the PPGs drops or the additional hardware overhead is not needed.
For a BS design with less bits accessible than required for processing per cycle, the respective part of the logic is kept idle as well.
Finally, the dataflow of the partial sums has to be determined.
Either partial sums are saved in individual registers and added outside of the PE (i.e., SA) to increase flexibility of the dataflow, or added in an adder tree inside of the PE (i.e., ST) to decrease the hardware overhead in form of registers.

We assumed that the optimal operand slice $k$ depends on multiple factors like the used precisions of the CNN or the available logic.
From a hardware perspective, powers-of-two could lead to an efficient solution, while not overinflating the design space.
The maximum investigated operand slice $k$ should allow the possibility to efficiently process more than one word-length option, with maximum weight word-length $w_\text{Q}$ being 8\thinspace bit and activation word-length always set to 8\thinspace bit to preserve accuracy \cite{Sze2019}. 
Hence, we investigate the operand slices of 1, 2, and 4\thinspace bit. 
\begin{figure}[htb]
\begin{center}
\includegraphics[scale= 0.65]{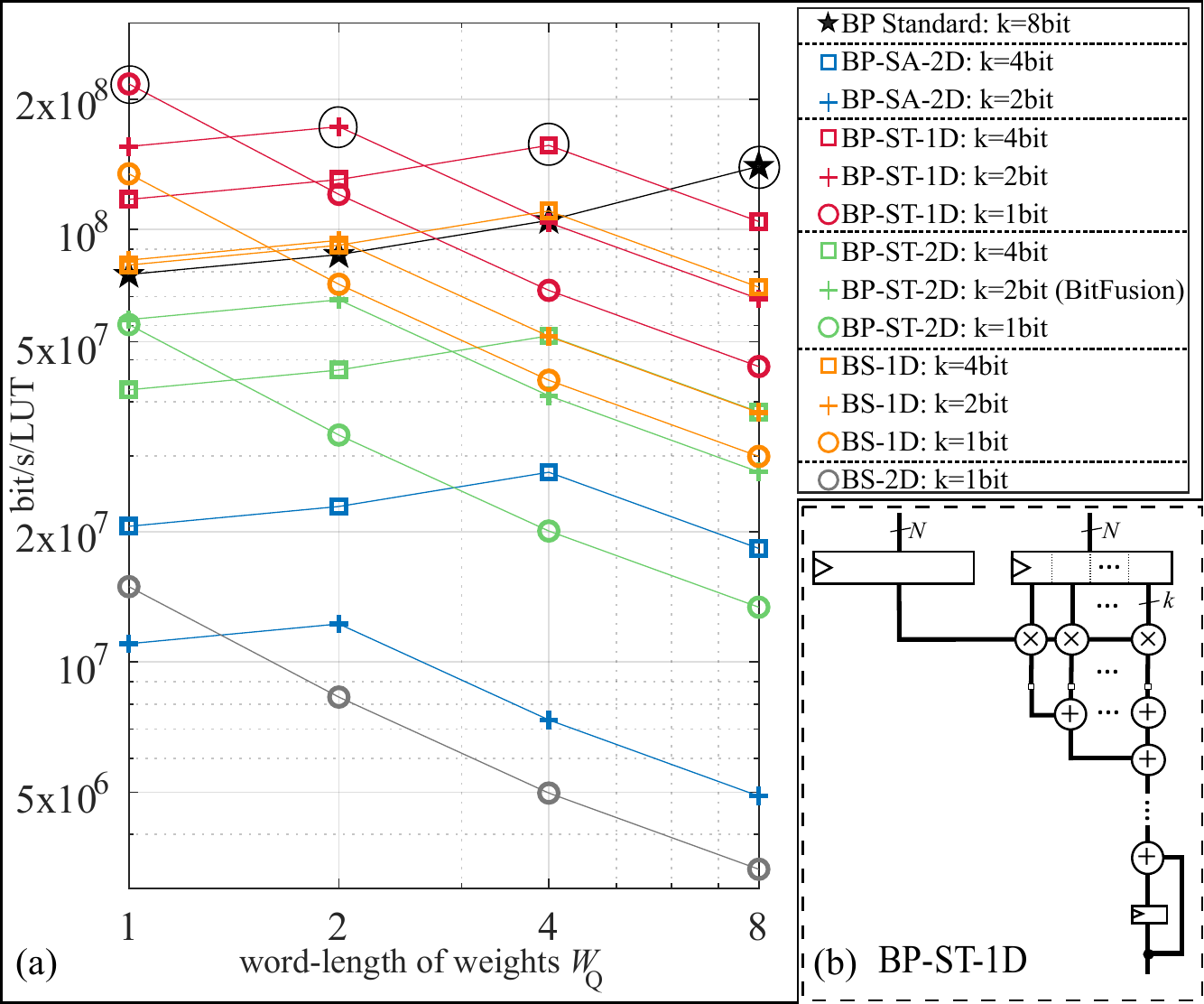}
\caption{Efficiency of PEs for processing activations of fixed 8\thinspace bit precision and weights of variable precision $w_\text{Q}$ in (a). Symbols represent the minimum precision $k$ supported by the design, i.e. $k$ bit\thinspace/\thinspace cycle for BS and operand slice $k$ for BP. The best performing PE design is encircled and shown in (b).}
\label{fig:PE_DSE}  
\end{center}
\vspace{-0.5cm}
\end{figure}

The quantitative comparison in Fig.\thinspace\ref{fig:PE_DSE} presents the results of the PE DSE (cf. blue box in Fig.\thinspace\ref{fig:DSE_flow_Chart}).
The dotted lines link related design points. 
Actual implementations are limited to a discrete word-length.
Typically, area efficiency is quantified as GOps\thinspace/\thinspace s\thinspace/\thinspace area, which translates to GOps\thinspace/\thinspace s\thinspace/\thinspace LUT in the case of an FPGA.
However, this metric does not take differences in word-length into account.
Hence, we use the processed bits\thinspace/\thinspace s\thinspace/\thinspace LUT
of a PE as quantitative optimization objective (maximization).
In Fig.\thinspace\ref{fig:PE_DSE}\thinspace a, the PE design with the best operand slice $k$ for a given weight word-length is encircled.
For all design points considering asymmetrical word-lengths, the Bit-Parallel Sum-Together one-dimensional-scaled design (BP-ST-1D) provides the best results. 
This is in agreement with the findings in \cite{Camus2019}.
Therefore, the chosen PE design is BP-ST-1D, cf. Fig.\thinspace\ref{fig:PE_DSE}\thinspace b.
\begin{figure}[htb]
\begin{center}
\includegraphics[scale = 0.65]{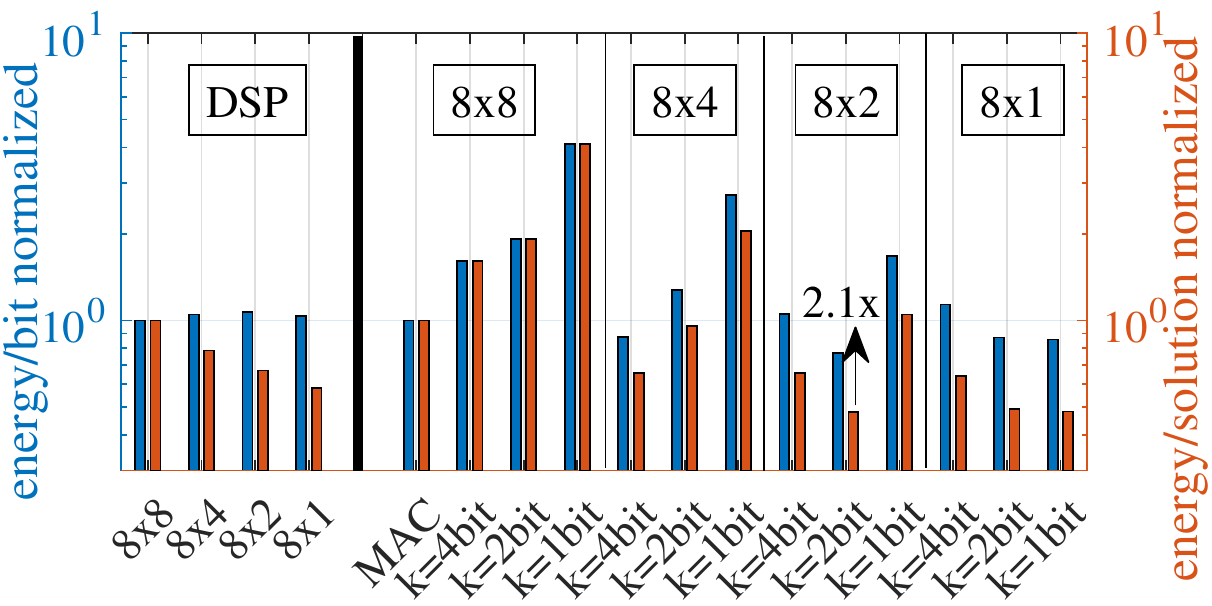}
\caption{Energy efficiency for BP-ST-1D in different operand slices $k$ normalized to 8\thinspace bit\thinspace$\times$\thinspace8\thinspace bit MAC.
To compare the effect of word-length scaling, the use of DSPs is normalized to the 8\thinspace bit\thinspace$\times$\thinspace8\thinspace bit DSP. The word-length is given in the boxes in bit as \text{activations}\thinspace$\times$\thinspace\text{weights}.
Bit normalized and solution normalized refers to the energy needed to process either one bit or one MAC solution, comprising all partial products. 
Both is normalized by the respective reference energy needed for processing either one bit or one final MAC.}
 \label{fig:PE_energy}  
\end{center}
\vspace{-0.5cm}
\end{figure}

Our gate-level-timing simulation indicated that the DSP hardmacros are 1.7$\times$
more energy efficient than the LUT-based PEs of identical word-length, while the former provides no or limited flexibility.
DSP and LUT-based PE designs are normalized in Fig.\thinspace\ref{fig:PE_energy} to their respective 8\thinspace bit\thinspace$\times$\thinspace8\thinspace bit reference.
The respective operand slice $k$ per PPG is indicated in bit.
Energy efficiency is maximized using slices that match the required word-length, e.g., comparing an 8\thinspace bit\thinspace$\times$\thinspace2\thinspace bit multiplication against a fixed-width 8\thinspace bit\thinspace$\times$\thinspace8\thinspace bit LUT-based operation provides an increase in energy efficiency of 2.1$\times$ (cf. Fig.\thinspace\ref{fig:PE_energy}).

As mentioned above, DSP resources are limited.
Taking the example of the Stratix\thinspace V GXA7, it features 256 DSPs.
LUT-based PEs provide between 2.7$\times$ and 7.8$\times$ more computational resources assuming word-lengths between 1\thinspace and  4\thinspace bit, thereby providing a proportionate increase in peak throughput.
As much as the flexibility in functionality supports more CNN variants (including layer- and channel-wise mixed-precision networks), it comes at the cost of additional control logic.
So, the final choice of the operand slice $k$ depends on the average word-length used in the adopted CNN.
Considering the reconfiguration opportunities using an FPGA platform, a dedicated image can be loaded that most optimally matches the specific CNN.
Hence, the results in Fig.\thinspace\ref{fig:accuracy_throughput} refer to individual FPGA images.

\subsection{Design Space Exploration of Dataflow}

The maximum feasible number of PEs was determined by assessing the PE architecture BP-ST-1D (cf. Fig.\thinspace\ref{fig:PE_DSE}\thinspace b) of different operand slices $k$, independently of the PE array dimensions.
This number of PEs serves as a threshold of PEs bound for the design space.
It is then searched for a PE array that offers the maximum Ops\thinspace/\thinspace Logic and Ops\thinspace/\thinspace Memory, while including hardware overhead and realizability with regards to the FPGA.
The choice of the operand slice $k$, as well as the PE array dimensions, are concurrently optimized considering the maximum number of realizable PEs and their respective average utilization when executing a given mixed-precision CNN. 
For this, all possible combinations in PE dimensions are automatically evaluated.
To reach highest throughput for each uniquely quantized CNN, the DSE in Fig.\thinspace\ref{fig:DSE_flow_Chart} has to be repeated, regarding the red and green box. 
As a result, a new FPGA accelerator design is created.
\begin{figure}[htb]
\begin{center}
\includegraphics[scale = 0.45]{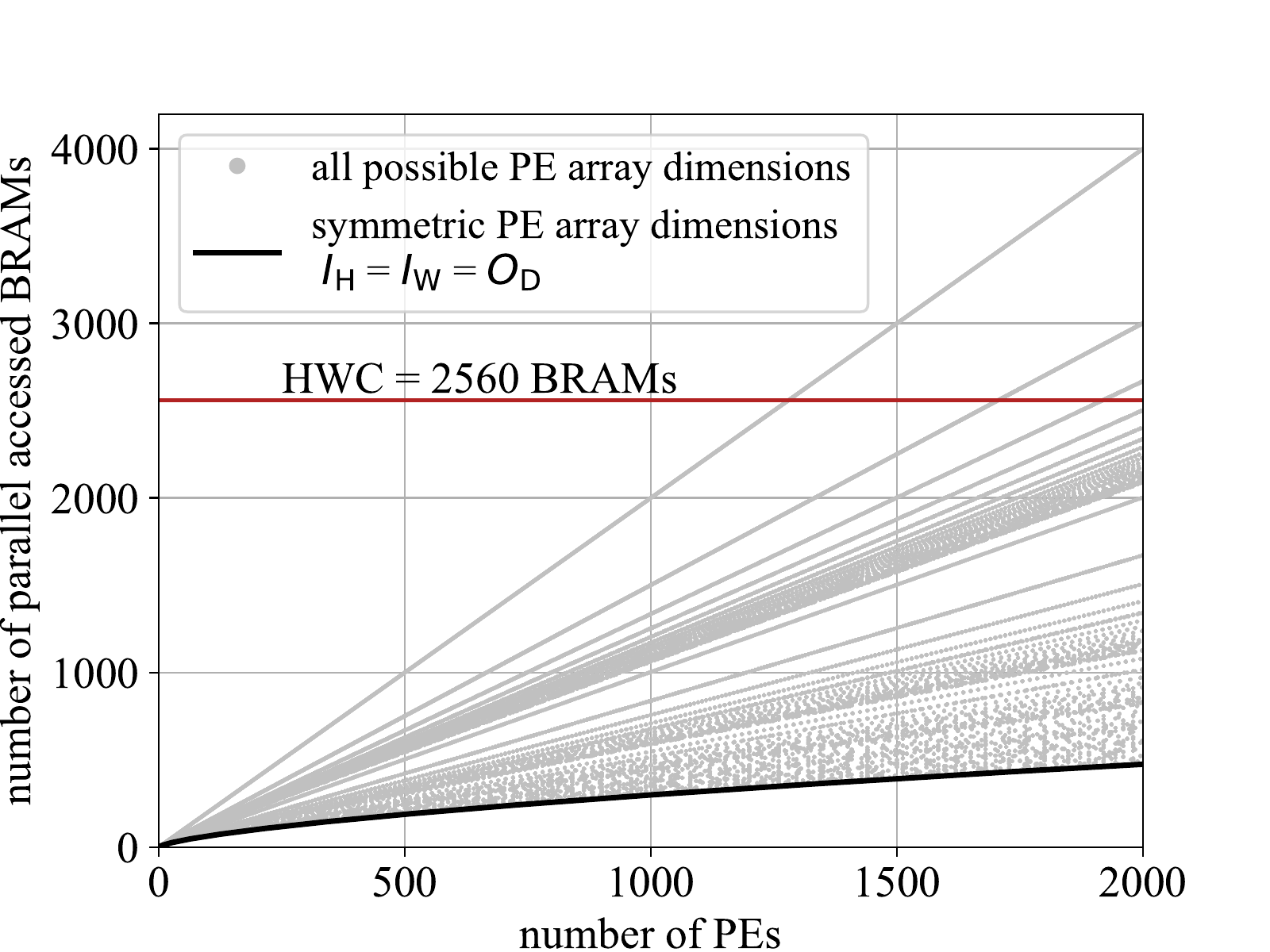}
\caption{{Impact of PE array dimensions on the necessary BRAMs for the resulting PE array. The operand slice $k$ equals 4\thinspace bit and all inputs (weights and activations) are 8\thinspace bit.}}
\label{fig:RAMvsPE}  
\end{center}
\vspace{-0.5cm}
\end{figure}

\begin{equation}
\begin{split}
\text{min}(BRAM_\text{NPA}) = 3 \times\sqrt[3]{N_\text{PE}^2} 
\\\ {\text{for } H = W = D \text{ and }N = w_\text{Q}}
\end{split}
\label{eq:N3}
\end{equation}
Our investigation shows that not only the total number of PEs is relevant for performance, but the PE array dimensions are as well.
The array dimensions dictate the average utilization of the implemented PEs and determine the possible spatial reuse of data fetched from the BRAM.
Combining Eq.\thinspace\ref{eq:NPE} and Eq.\thinspace\ref{eq:reuse} for a symmetric PE array, Eq.\thinspace\ref{eq:N3} shows the minimum $BRAM_\text{NPA}$.
The total number of parallel BRAM accesses is lower for symmetric PE array dimensions (height $H$, width $W$, and depth $D$ of the PE array are the same) than for asymmetric PE array dimensions, cf. Fig.\thinspace\ref{fig:RAMvsPE}.
Besides, if the number of BRAMs decreases, the hardware overhead declines.
Hence, the data points with less parallel BRAM accesses are preferable for implementation.
Nevertheless, the most optimal PE dimensions with regards to the maximization of Ops\thinspace/\thinspace resources (cf. Fig.\thinspace\ref{fig:DSE_flow_Chart} red box) were surprisingly not symmetrical (cf. Table\thinspace\ref{tab:PEdimension}). 
Because the PE array utilization highly depends on the CNN (cf. Eq.\thinspace\ref{eq:util}), which typically does not provide symmetrical dimensions $I_\text{H}$, $I_\text{W}$, and $O_\text{D}$.
\begin{table} [htbp]
\addtolength{\tabcolsep}{-3.5pt}
\caption{{Chosen PE array dimensions}}
\begin{center}
\begin{tabular}{|c|c|c|c|c|c|}
\hline
\textbf{CNN model} & \textbf{operand slice} $k$ & \multicolumn{3}{|c|}{\textbf{PE array dimensions}}&\textbf{$N$}$_\textbf{PE}$ \\
\cline{2-5}
 &  (bit)&  $H$&  $W$&  $D$&\\
\hline
\hline
ResNet-18       & 1 &  7&3 &32& 672 \\
\cline{2-6}
                &2 &  7&5&37& 1295\\
\cline{2-6}
                &4&  7&4&66& 1848 \\
\hline
ResNet-50 \&       & 1 &  7&3 &33 &693 \\
\cline{2-6}
ResNet-152         &2 &  7&5&37&1295\\
\cline{2-6}
                &4&  7&4&71 &1988\\
\hline
\end{tabular}
\label{tab:PEdimension}  
\end{center}
\vspace{-0.5cm}
\end{table}

\subsection{System Level Performance}

\begin{figure}[htb]
\begin{center}
\includegraphics[scale = 0.62]{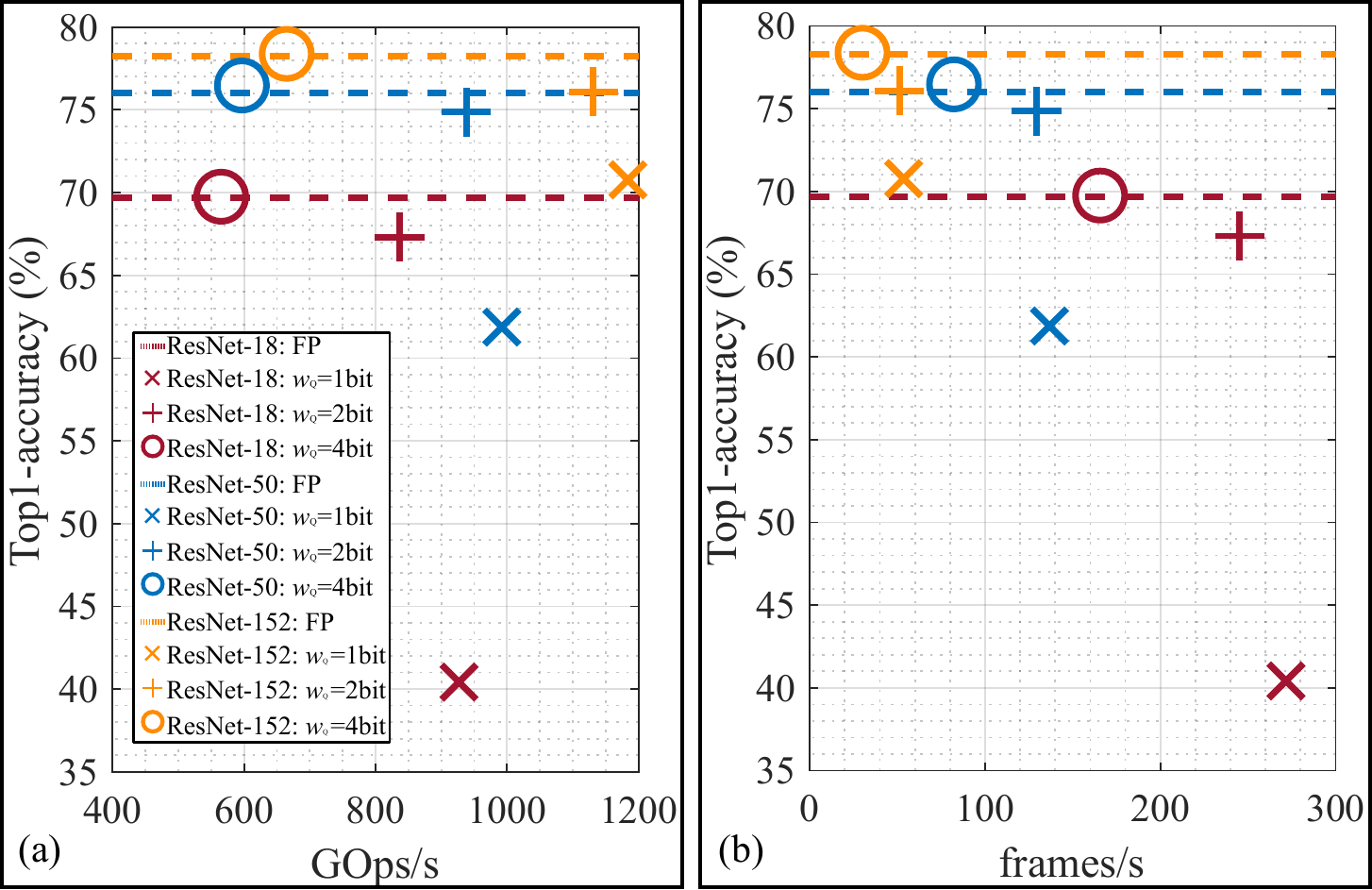}
\caption{ResNet-\# (18, 50, 152) accuracy (\%) versus performance of our FPGA design with operand slice $k$ = $w_\text{Q}$. For the floating-point ResNets only the respective accuracies are shown as FP.}
\label{fig:accuracy_throughput}
\vspace{-0.5cm}
\end{center}
\end{figure}

Our FPGA designs support numerous mixed-precision CNNs and translates the word-length reduction directly into throughput gains. 
Their accuracy-throughput trade-off is shown in Fig.\thinspace\ref{fig:accuracy_throughput}.
It shows the benefit in accuracy for deeper CNN architectures as well as of higher word-lengths.
For CNNs in this work, we fix activations as well as first and last layer weights to 8bit. The word-length of all other weights is set to $w_\text{Q}$.
For inference, Eq.\thinspace\ref{eq:quant} quantizes activations and weights, here both represented with $\nu$, with the round to nearest function $\lfloor\cdot\rceil$ and saturated with the clamp function to its upper bound Q$_\text{p}$ and to its lower bound Q$_\text{n}$.
\textcolor{black}{Activations are quantized in an unsigned fashion, hence, for activations $Q_\text{n}$ = 0 and $Q_\text{p}$ = $2^{b}-1$.
Weights are quantized as signed numbers, so, for weights $Q_\text{n}$ = $-2^{b-1}$ and $Q_\text{p}$ = $2^{b-1}-1$.}
FP represents the floating-point value, quant represents the quantized value, and int represents the integer value.
The quantizer step size $\gamma$ is initialized and trained as in \cite{Esser2019} described.
By quantization aware training of 30 epochs of pretrained ResNets from Torchvision, the floating-point weights are updated.
For FP, all weights and activations are kept in 32\thinspace bit floating-point.
Mixed-precision CNNs can surpass floating-point accuracy, as indicated in Fig.\thinspace\ref{fig:accuracy_throughput} for ResNet-18, ResNet-50 and ResNet-152 with $w_\text{Q}$ equals 4\thinspace bit.
By further optimizing the word-length per layer, the accuracy-memory footprint trade-off (cf. Table\thinspace\ref{tab:Resnets}) could be even more pronounced, as shown in \cite{Yang2019, Wu2018, Wang2018, Esser2019, Cho2020, Wang2020}.
We leave the optimization of word-length per layer to future work, because we focus in this work on the hardware accelerator. Those future optimizations could improve accuracy while maintaining memory footprint.
\begin{equation}
\begin{split}
{\nu_\text{quant} = \nu_\text{int} \times \gamma}
\\\ {{\text{with }\nu_\text{int} = \lfloor(\text{clamp}(\frac{\nu_\text{FP}}{\gamma},Q_\text{n} ,Q_\text{p} ))\rceil}}
\end{split}
\label{eq:quant}
\end{equation}

\begin{table} [htbp]
\addtolength{\tabcolsep}{-3.5pt}
\caption{Accuracy versus memory footprint}
\begin{center}
\begin{tabular}{|c|c|c|c|c|c|}
\hline
\textbf{CNN} &\textbf{$w$}$_\text{\textbf{Q}}$ &\textbf{memory footprint}&\textbf{compression }& \multicolumn{2}{|c|}{\textbf{accuracy}} \\
\cline{2-6}
                         &(bit)& (MB)&\textbf{factor}&Top-1&Top-5\\
\hline
\hline
ResNet-18       &FP&352&1.0&69.69&89.07 \\
\cline{2-6}
                &1&{69}&5.1&40.42&65.29 \\
\cline{2-6}
                &2&72&4.9& 67.31&87.48 \\
\cline{2-6}
                &4&77&4.6&69.75&89.10 \\
\hline
\hline
ResNet-50       &FP&662&1.0 &76.00&92.93\\
\cline{2-6}
                &1&111& 6.0 &61.87 &83.95\\
\cline{2-6}
                &2&118& 5.6 &74.86& 92.24\\
\cline{2-6}
                &4&134&4.9  &76.47&93.07 \\
\hline
\hline
ResNet-152      &FP&1767& 1.0 &78.26&93.94\\
\cline{2-6}
                &1&145& 12.2 & 70.77&90.02\\
\cline{2-6}
                &2&188& 9.4 & 76.09&92.90\\
\cline{2-6}
                &4&272& 6.5 & 78.38&94.00\\
\hline
\end{tabular}
\label{tab:Resnets}  
\end{center}
\vspace{-0.4cm}
\end{table}
Table\thinspace\ref{tab:8x8bit} presents the energy\thinspace/\thinspace frame for four differently quantized ResNet-18 on three different accelerator designs with one image per batch.
The accelerator designs are optimized for ResNet-18 and PEs of the type BP-ST-1D with either operand slice $k$ of 1, 2, or 4\thinspace bit.
The contributions to the energy figures in Table\thinspace\ref{tab:8x8bit} are based on DDR3 accesses with 70\thinspace pJ\thinspace/\thinspace bit \cite{Malladi2012}, a BRAM memory model equivalent to the Stratix\thinspace V M20K memory block as well as the energy for processing estimated by the available Quartus toolchain for Stratix\thinspace IV and scaling to Stratix\thinspace V using results of \cite{Latotzke2021}.

Table\thinspace\ref{tab:8x8bit} highlights the impact of operand slices.
Supporting weights of 8\thinspace bit, a PE array with 2\thinspace bit operand slices features less energy for BRAM accesses as compared to an array with 1\thinspace bit operand slices. 
This is due to an increased reuse of data fetched from the BRAM in the first case. A larger PE array contributes to more data reuse options.
The energy for BRAM accesses is dominated by the partial sum with 30\thinspace bit.

Besides, Table\thinspace\ref{tab:8x8bit} shows also the impact of operand slice $k$ as function of the predominant word-length used in a given CNN. 
If the operand slice matches the average word-length of a CNN, the total energy\thinspace/\thinspace frame is minimized for the 1\thinspace bit case.
Then again, a CNN with no binary weights does not benefit from the PE design, hence it uses {6.36$\times$} more energy than the CNN with mostly 1\thinspace bit weights.
The factor of total energy reduction between the 4\thinspace bit case and the 2\thinspace bit case is 1.34$\times$, but the factor of total energy reduction between the 2\thinspace bit case and the 1\thinspace bit case is only 1.02$\times$.
This is caused by the high efficiency of the PPG with 2\thinspace bit operand slice (cf. Fig\thinspace \ref{fig:PE_energy}).

To conclude, if the word-length is smaller than the operand slice, PPGs are not fully utilized.
Furthermore, a smaller operand slice reduces the total energy\thinspace/\thinspace frame for input word-lengths of same size.
Then again, higher operand slices reduce the shift logic and decrease the size of the adder tree, so more PEs can be implemented with the same resources.
Hence, the peak throughput increases.
\begin{table}[htbp]
\addtolength{\tabcolsep}{-3pt}
\caption{{Impact of operand slices processing ResNet-18}}
\begin{center}
\begin{tabular}{|c|c|c|c|c|c|c|}
\hline
operand slice $k$ (bit)&\textbf{1}&\textbf{2}&\textbf{4}&\textbf{1}&\textbf{2}&\textbf{4}  \\
\hline
\hline
inner layers $w_\text{Q}$ (bit)&\multicolumn{3}{|c|}{8}&1&2&4  \\
\hline
Top1 accuracy (\%)&\multicolumn{3}{|c|}{70.40}&40.42&67.31&69.75\\
\hline
Top5 accuracy (\%)&\multicolumn{3}{|c|}{89.62}&65.29&87.48&89.10\\
\hline
\hline
kLUTs&392.24&327.68 &243.94 &380.35 &331.52 &243.94  \\
\hline
BRAM&\multicolumn{3}{|c|}{{2470}}&{1644}&{1762}&{1998}\\
\hline
frequency (MHz)&124&127&96&124&127&96\\
\hline
\hline
energy\thinspace/\thinspace frame for &&&&&&\\
computation (mJ)& 100.90&47.06& 23.40  &11.80  &11.76 &16.06 \\
\hline
energy\thinspace/\thinspace frame for &&&&&&\\
BRAM accesses (mJ)& {7.59}&{5.42}&{5.85}&{1.35}&{1.55}&{3.21} \\
\hline
energy\thinspace/\thinspace frame for &&&&&&\\
DDR3 accesses (mJ)& 6.24& 6.24& 6.24 &4.90&5.10& 5.48\\ 
\hline
total energy\thinspace/\thinspace frame (mJ)&114.73 & 58.72& 35.49 &18.05 & 18.41  & 24.75   \\
\hline
\hline
frames/s&46.86&83.81&97.25  &271.68&245.23&165.63 \\
\hline
GOps/s&159.87&285.94& 331.77 &926.84  &836.61 & 565.05 \\
\hline
GOps/s/W&0.066&0.418&0.930&14.142&11.320&3.831 \\
\hline
\end{tabular}
\label{tab:8x8bit}
\end{center}
\vspace{-0.1cm}
\end{table}

Table\thinspace \ref{tab:sota} provides a comparison of architectures supporting layer-wise mixed-precision CNNs processing ImageNet.
Please note, only reference \cite{Ma2018} uses an identical FPGA. 
In any case, none reports energy numbers.
Our designs abstain from using DSPs given their limited number (256 on Stratix\thinspace V).
Furthermore, only the work of \cite{Nguyen2021}, \cite{Maki2018}, and our work support channel-wise mixed-precision CNNs.
But \cite{Nguyen2021} is limited to binary and 8\thinspace bit, whereas our implementation supports 1, 2, 4, and 8\thinspace bit, resulting in better accuracy-throughput trade-offs.
As a common practice, we count one MAC operation as two operations.
For references, that do not follow this practice, Ops numbers are multiplied by a factor of 2.

Our work outperforms for ResNet-152 with 1.13\thinspace TOps\thinspace/\thinspace s by 1.56$\times$ Nguyen et al. \cite{Nguyen2021} and by 4.09$\times$ Ma et al. \cite{Ma2018}, and it outperforms for ResNet-50 with 938\thinspace GOps\thinspace/\thinspace s Maki et al. \cite{Maki2018} by 9.84$\times$.
Embracing logic-based MAC realizations as in \cite{Maki2018}, PE count is increased by 2.63$\times$ (for ResNet-18 with $k$ = 1\thinspace bit) up to 7.77$\times$ (for ResNet-152 with $k$ = 4\thinspace bit) when compared to the 256 DSPs on a Stratix\thinspace V (cf. Table\thinspace\ref{tab:PEdimension}).
This increase is visible in the total PE array dimensions and, therefore, in the total throughput as well as data reuse and with this in the total BRAM access reduction.

\begin{table}[htbp]
\addtolength{\tabcolsep}{-4.5pt}
\caption{{State-of-the-art comparison}}
\begin{center}
\begin{tabular}{|c|c|c|c|c|ccc|}
\hline
\textbf{Metrics} & \textbf{\cite{Blott2018}} & \textbf{\cite{Maki2018}} & \textbf{\cite{Ma2018}} & \textbf{\cite{Nguyen2021}} & \multicolumn{3}{c|}{\textbf{this work}}                     \\\hline\hline
                 & DoReFa     & ResNet     & ResNet     & ResNet      & \multicolumn{1}{c|}{ResNet} & \multicolumn{1}{c|}{ResNet} & ResNet \\
CNN              & Net/PF     & -50        & -152       & -152        & \multicolumn{1}{c|}{-50}    & \multicolumn{1}{c|}{-152}   & -152   \\ \hline
Top1$^{\mathrm{a}}$           &            &            &             &             & \multicolumn{1}{c|}{}       & \multicolumn{1}{c|}{}       &        \\
accuracy         & 50.3       & -          & -          & -           & \multicolumn{1}{c|}{74.86}  & \multicolumn{1}{c|}{76.09}  &  78.17      \\ \hline
Top5$^{\mathrm{a}}$           &             &             &           &             & \multicolumn{1}{c|}{}       & \multicolumn{1}{c|}{}      &        \\
accuracy         & 74.0       & 91.9       & -          & -           & \multicolumn{1}{c|}{92.24}  & \multicolumn{1}{c|}{92.90}  & 93.96        \\ \hline
word-length      &            &            &            &             & \multicolumn{2}{c|}{}                                     &        \\
weights          & 1          & 1-16       & 16         & 8$^{\mathrm{d}}$          & \multicolumn{2}{c|}{2$^{\mathrm{b}}$}                                  & 8      \\ \hline
word-lenght      &            &            &            &             & \multicolumn{3}{c|}{}                                              \\
activations      & 2          & 8          & 16         & 8$^{\mathrm{d}}$& \multicolumn{3}{c|}{8}                                             \\ \hline \hline
FPGA             & PYNQ-Z1    & ZCU\thinspace102   &  Stratix\thinspace V  & Virtex\thinspace7    & \multicolumn{3}{c|}{Stratix\thinspace V} \\ \hline
node (nm)        & 28         & 16         & 28         & 28          & \multicolumn{3}{c|}{28}                                            \\ \hline
f (MHz)          & 100        & 100        & 150        & 200         & \multicolumn{3}{c|}{127}                                           \\ \hline
operand slice:   & 2\thinspace$\times$\thinspace1&8\thinspace$\times$\thinspace1&16\thinspace$\times$\thinspace16&8\thinspace$\times$\thinspace1 \&& \multicolumn{3}{c|}{8 x 2}\\
$N$\thinspace bit\thinspace$\times$\thinspace $k$\thinspace bit && & &8\thinspace$\times$\thinspace8 $^{\mathrm{c}}$ &\multicolumn{3}{c|}{{}} \\\hline
BRAMs             & 278        &    900     &  2385      &     716     & \multicolumn{2}{c|}{1762}                                 & 2470   \\ \hline
DSPs              &-           &     0      &    256     &       2515  & \multicolumn{3}{c|}{0}                                             \\ \hline
kLUT             &35.7        &     57     &    370     &   280.4     & \multicolumn{3}{c|}{331.5}                                         \\ \hline
BRAM in \%       & 99         &    98      &  93        &     69      & \multicolumn{2}{c|}{69}                                   &   96    \\ \hline
DSPs in \%       & -          &     0      &    100     & 90          & \multicolumn{3}{c|}{0}                                             \\ \hline
LUT in \%        & 67         &21          &78          &  92         & \multicolumn{3}{c|}{71}                                             \\ \hline
channel 
$^{\mathrm{e}}$  & no        & yes        &no          &   yes       & \multicolumn{3}{c|}{yes}                                           \\ \hline
flexible 
 $^{\mathrm{f}}$ &  yes       & yes        &    no      &   yes       & \multicolumn{3}{c|}{yes}                                           \\ \hline
CONV only        &  no        &yes         &    no      &  yes        & \multicolumn{3}{c|}{yes}                                           \\ \hline
GOps\thinspace/\thinspace s
                 &258$^{\mathrm{g}}$&95.4$^{\mathrm{g}}$&    276.6   &726.0        & \multicolumn{1}{c|}{938.33} & \multicolumn{1}{c|}{1131.38} & 311.16      \\ \hline
frames\thinspace/\thinspace s
                 &  -         & -          &  12.23     & 32.1        & \multicolumn{1}{c|}{129.38} & \multicolumn{1}{c|}{51.19}      & 14.08      \\ \hline
mJ\thinspace/\thinspace frame
                 &  -         & -          &-           &-            & \multicolumn{1}{c|}{36.56}  & \multicolumn{1}{c|}{97.71}  & 367.69  \\ \hline
GOps\thinspace/\thinspace s\thinspace/\thinspace W
                 &  102$^{\mathrm{g}}$        & -          &-           & -           & \multicolumn{1}{c|}{198.39} & \multicolumn{1}{c|}{226.20} & 60.11   \\ \hline
\multicolumn{8}{c}{$^{\mathrm{a}}$ ImageNet $^{\mathrm{b}}$ first and last layer are 8\thinspace bit $^{\mathrm{c}}$ equivalent to PE size}\\
\multicolumn{8}{c}{$^{\mathrm{d}}$ Folding two 8x8 mult per 16bit DSP to achieve f$_\text{clk}$$\times$DSPs$\times$2$\times$efficiency }\\
\multicolumn{8}{c}{= 200MHz $\times$ 2515 $\times$ 2 $\times$ 72.2$\%$  = 726GOps}\\
\multicolumn{8}{c}{$^{\mathrm{e}}$ supports channel-wise mixed-precision}\\
\multicolumn{8}{c}{$^{\mathrm{f}}$ unknown input word-length can be processed}\\
\multicolumn{8}{c}{$^{\mathrm{g}}$ Ops\thinspace/\thinspace s in reference refer to MACs\thinspace/\thinspace s, so the value is multiplied by two}\\
\end{tabular}
\label{tab:sota}
\end{center}
\vspace{-0.1cm}
\end{table}

\section{Conclusion}
\label{sec:Conclusion}

The design space of mixed-precision CNN hardware accelerators is enormous.
In a structured approach, the flexibility-throughput-energy trade-off was quantitatively assessed along the design dimension on the bit-level, PE-level, and architectural level.
Considering the granularity of natively supported word-lengths from 1 to 8\thinspace bit, a dedicated optimum exists as a function of the distribution of word-lengths in the targeted CNN model.
A reduction in energy up to {6.36$\times$} is reached when comparing a mixed-precision CNN against a CNN with fixed word-length of 8\thinspace bit.
This work enhances flexibility by supporting processing varying precision of a given CNN as well as taking advantage of the FPGA platform to adjust the accelerator design in accordance to a specific CNN for maximum hardware utilization, and hence, to achieve maximum throughput.
Compared to state-of-the-art in mixed-precision FPGA accelerators, this work increases its frames\thinspace/\thinspace s by 1.56$\times$ for ResNet-152 compared to \cite{Nguyen2021} and achieves 9.84$\times$ more GOp\thinspace/\thinspace s for ResNet-50 compared to \cite{Maki2018}.


\begin{thebibliography}{00}
\bibitem{Horowitz2014} M. Horowitz, ``1.1 computing's energy problem (and what we can do about it),'' In Proceedings of the IEEE International Solid-State Circuits Conference (ISSCC), 2014, pp. 10--14.

\bibitem{Stadtmann2020} T. Stadtmann, C. Latotzke, and T. Gemmeke, ``From Quantitative Analysis to Synthesis of Efficient Binary Neural Networks,'' In Proceedings of the 19th IEEE International Conference On Machine Learning And Applications (ICMLA), 2020, pp. 93--100.

\bibitem{ImageNet} J. Deng, W. Dong, R. Socher, L. J. Li, K. Li, and L. Fei-Fei ``Imagenet: A large-scale hierarchical image database,'' In Proceedings of the 2009 IEEE Conference on Computer Vision and Pattern Recognition (CVPR), 2009, pp. 248--255.

\bibitem{Sze2019}  Y. H. Chen, T. J. Yang, J. S. Emer, and V. Sze, ``Eyeriss v2: A Flexible Accelerator for Emerging Deep Neural Networks on Mobile Devices,'' In IEEE Journal on Emerging and Selected Topics in Circuits and Systems (IEEE J. Emerg. Sel), vol. 9, no. 2, 2019, pp. 292--308.

\bibitem{Yang2019} J. Yang, X. Shen, J. Xing, X. Tian, H. Li, B. Deng, J. Huang, and X. S. Hua, ``Quantization Networks,'' In Proceedings of the IEEE Conference on Computer Vision and Pattern Recognition (CVPR), 2019, pp. 7308--7316.

\bibitem{Wu2018} B. Wu, Y. Wang, P. Zhang, Y. Tian, P. Vajda, and K. Keutzer, ``Mixed precision quantization of convnets via differentiable neural architecture search,'' arXiv preprint, 2018, arXiv:1812.00090.

\bibitem{Wang2018} K. Wang, Z. Liu, Y. Lin, J. Lin, and S. Han, ``Haq: Hardware-aware automated quantization with mixed precision,'' In IEEE Conference on Computer Vision and Pattern Recognition (CVPR), 2019, pp. 8612--8620.

\bibitem{Cho2020} S. Cho and S. Yoo, ``Per-channel quantization level allocation for quantizing convolutional neural networks,'' In Proceedings of the IEEE International Conference on Consumer Electronics-Asia (ICCE-Asia), 2020, pp. 1--3.

\bibitem{Wang2020} T. Wang, K. Wang, H. Cai, J. Lin, Z. Liu, H. Wang, Y. Lin, and S. Han, ``Apq: Joint search for network architecture, pruning and quantization policy,'' In Proceedings of the 2020 IEEE\thinspace/\thinspace CVF Conference on Computer Vision and Pattern Recognition (CVPR), 2020, pp. 2078--2087.

\bibitem{Esser2019} S. Esser, J. McKinstry, D. Bablani, R. Appuswamy, and D. Modha, ``Learned step size quantization,'' arXiv preprint, 2019, arXiv:1902.08153.

\bibitem{Lin2016} D. Lin, S. Talathi, and S. Annapureddy,``Fixed point quantization of deep convolutional networks," In Proceedings of the 33rd International Conference on International Conference on Machine Learning (ICML), 2016, pp. 2849--2858.

\bibitem{Mittal2020} S. Mittal, "A survey of FPGA-based accelerators for convolutional neural networks," In Neural Computing and Applications (Neural. Comput. Appl.), vol. 32, no. 4, 2020, pp. 1109--1139.

\bibitem{Latotzke2021} C. Latotzke, and T. Gemmeke, ``Efficiency Versus Accuracy: A Review of design Techniques for DNN Hardware Accelerators,'' In IEEE Access, vol. 9, 2021, pp. 9785--9799.

\bibitem{ResNet} K. He, X. Zhang, S. Ren, and J. Sun, ``Deep residual learning for image recognition,'' In Proceedings of the 2016 IEEE Conference on Computer Vision and Pattern recognition, (CVPR), 2016, pp. 770--778.

\bibitem{Ma2018} Y. Ma, Y. Cao, S. Vrudhula, and J. Seo, ``Optimizing the convolution operation to accelerate deep neural networks on FPGA,'' In IEEE Transactions on Very Large Scale Integration (IEEE VLSI Systems), vol. 26, no. 7, 2018, pp. 1354--1367.

\bibitem{Tsimpourlas2018} F. Tsimpourlas, L. Papadopoulos, A. Bartsokas, and D. Soudris, ``A design space exploration framework for convolutional neural networks implemented on edge devices,'' In IEEE Transactions on Computer-Aided Design of Integrated Circuits and Systems (IEEE TCAD), vol. 37, no. 11, 2018, pp. 2212--2221.

\bibitem{Parashar2019} A. Parashar, P. Raina, Y. Shao, Y. Chen, V. Ying, A. Mukkara, R. Venkatesan, B. Khailany, S. Keckler, and J. Emer, ``Timeloop: A systematic approach to dnn accelerator evaluation,'' In Proceedings of the 2019 IEEE International Symposium on Performance Analysis of Systems and Software (IEEE ISPASS), 2019, pp. 304--315.

\bibitem{Yang2020} X. Yang, M. Gao, Q. Liu, J. Setter, J. Pu, A. Nayak, S. Bell, K. Cao, H. Ha, P, Raina, and C. Kozyrakis, ``Interstellar: Using halide's scheduling language to analyze dnn accelerators,'' In Proceedings of the 25th International Conference on Architectural Support for Programming Languages and Operating Systems (ASPLOS), 2020, pp. 369--383.

\bibitem{Rahman2017} A. Rahman, S. Oh, J. Lee, and K. Choi, ``Design space exploration of FPGA accelerators for convolutional neural networks,'' In Proceedings of the 2017 Design, Automation \& Test in Europe Conference (DATE), 2017, pp. 1147--1152.

\bibitem{Reggiani2019} E. Reggiani, M. Rabozzi, A. Nestorov, A. Scolari, L. Stornaiuolo, and M. Santambrogio, ``Pareto optimal design space exploration for accelerated CNN on FPGA,'' In Proceedings of the 2019 IEEE International Parallel and Distributed Processing Symposium Workshops (IEEE IPDPSW), 2019, pp. 107--114.

\bibitem{Zhong2017} G. Zhong, A. Prakash, S. Wang, Y. Liang, T. Mitra, and S. Niar, 
``Design Space exploration of FPGA-based accelerators with multi-level parallelism,'' In Proceedings of the 2017 Design, Automation \& Test in Europe Conference (DATE), 2017, pp. 1141--1146. 

\bibitem{Lu2017} L. Lu, Y. Liang, Q. Xiao, and S. Yan, ``Evaluating fast algorithms for convolutional neural networks on FPGAs,'' In Proceedings of the IEEE 25th Annual International Symposium on Field-Programmable Custom Computing Machines (IEEE FCCM), 2017, pp. 101--108.

\bibitem{Veneries2017} S. Venieris, and C. Bouganis, ``Latency-driven design for FPGA-based convolutional neural networks,'' In Proceedings of the 27th International Conference on Field Programmable Logic and Applications (FPL), 2017, pp. 1--8. 

\bibitem{Kwon2020} H. Kwon, P. Chatarasi, V. Sarkar, T. Krishna, M. Pellauer, and A. Parashar, 
``Maestro: A data-centric approach to understand reuse, performance, and hardware cost of dnn mappings,'' In IEEE MICRO, vol. 40, no. 3, 2020, pp. 20--29.

\bibitem{Samajadar2019} A. Samajdar, Y. Zhu, P. Whatmough, M. Mattina, and T. Krishna, ``Scale-sim: Systolic cnn accelerator simulator,'' arXiv preprint, 2018, arxiv:1811.02883.

\bibitem{Blott2018} M. Blott, T. Preußer, N. Fraser, G. Gambardella, K. O’Brien, Y. Umuroglu, M. Leeser, and K. Vissers, ``FINN-R: An end-to-end deep-learning framework for fast exploration of quantized neural networks,'' In ACM Transactions on Reconfigurable Technology and Systems (ACM TRETS), vol. 11, no. 3, 2018, pp. 1--23.

\bibitem{Nguyen2021} D. Nguyen, H. Kim, and H. Lee, ``Layer-specific optimization for mixed data flow with mixed-precision in FPGA design for CNN-based object detectors,'' In IEEE Transactions on Circuits and Systems for Video Technology (IEEE Trans. Circuits Syst. Video Technol.), vol. 31, no. 6, 2020, pp. 2450--2464.

\bibitem{Sharma2018} H. Sharma, J. Park, N. Suda, L. Lai, B. Chau, V. Chandra, and H. Esmaeilzadeh,
``Bit fusion: Bit-level dynamically composable architecture for accelerating deep neural network,'' In Proceedings of the ACM\thinspace/\thinspace IEEE 45th Annual International Symposium on Computer Architecture (ISCA), 2018, pp. 764--775.

\bibitem{Ryu2019} S. Ryu, H. Kim, W. Yi, and J. Kim, ``Bitblade: Area and energy-efficient precision-scalable neural network accelerator with bitwise summation,'' In Proceedings of 56th Annual Design Automation Conference (DAC), 2019, pp. 1--6.

\bibitem{Camus2019} V. Camus, L. Mei, C. Enz, and M. Verhelst, ``Review and benchmarking of precision-scalable multiply-accumulate unit architectures for embedded neural-network processing,'' In IEEE Journal on Emerging and Selected Topics in Circuits and Systems (IEEE J. Emerg. Sel), vol. 9, no. 4, 2019, pp. 697--711.

\bibitem{Kim2019} H. Kim, Q. Chen, T. Yoo, T. T. H. Kim, and B. Kim, ``A 1-16b precision reconfigurable digital in-memory computing macro featuring column-mac architecture and bit-serial computation,'' In Proceedings of the IEEE 45th European Solid State Circuits Conference (ESSCIRC), 2019, pp. 345--348.

\bibitem{Williams2009} S. Williams, A. Waterman, and D. Patterson. "Roofline: an insightful visual performance model for multicore architectures." In Communications of the ACM vol. 52, no. 4, 2009, pp. 65--76.

\bibitem{Malladi2012} K. Malladi, F. Nothaft, K. Periyathambi, B. Lee, C. Kozyrakis, and M. Horowitz, 
``Towards energy-proportional datacenter memory with mobile DRAM,'' In Proceedings of the 39th Annual International Symposium on Computer Architecture (ISCA), 2012, pp. 37--48.

\bibitem{Maki2018} A. Maki, D. Miyashita, K. Nakata, F. Tachibana, T. Suzuki, and J. Deguchi,
``Fpga-based cnn processor with filter-wise-optimized bit precision,'' In Proceedings of the 2018 IEEE Asian Solid-State Circuits Conference (A-SSCC), 2018, pp. 47--50.
\end{thebibliography}
\end{document}